\documentclass[journal=ancac3,manuscript=article]{achemso}
\usepackage{multicol}

\usepackage{bm}
\usepackage{graphicx}
\usepackage{subfig}
\usepackage[]{amsmath, amssymb} % Formula subscripts using \ce{}
\usepackage{soul}

\newcommand{\cmmnt}[1]{}
\usepackage{color,soul}
\usepackage{comment}
\usepackage{xcolor}

\usepackage{lineno}                                                                     \usepackage{marginnote}

\mciteErrorOnUnknownfalse

\author{Maria G. Burdanova}
\affiliation{University of Warwick, Department of Physics, Gibbet Hill Road, Coventry, CV4 7AL, United Kingdom.}
\alsoaffiliation{Laboratory of Nanooptics and Plasmonics, Moscow Institute of Physics and Technology, 9, Institutsky Lane, Dolgoprudny, 141700, Russian Federation.}

\author{Ming Liu}
\affiliation{Department of Mechanical Engineering, The University of Tokyo, Tokyo 113-8656, Japan.}

\author{Michael Staniforth}
\affiliation{University of Warwick, Department of Physics, Gibbet Hill Road, Coventry, CV4 7AL, United Kingdom.}

\author{Yongjia Zheng}
\affiliation{Department of Mechanical Engineering, The University of Tokyo, Tokyo 113-8656, Japan.}

\author{Rong Xiang}
\affiliation{Department of Mechanical Engineering, The University of Tokyo, Tokyo 113-8656, Japan.}

\author{Shohei Chiashi}
\affiliation{Department of Mechanical Engineering, The University of Tokyo, Tokyo 113-8656, Japan.}

\author{Anton Anisimov}
\affiliation{Canatu Ltd., Helsinki FI‐00390, Finland.}

\author{Esko I. Kauppinen}
\affiliation{Department of Applied Physics, Aalto University School of Science, Espoo 15100, FI-00076 Aalto, Finland.}

\author{Shigeo Maruyama}
\affiliation{Department of Mechanical Engineering, The University of Tokyo, Tokyo 113-8656, Japan.}

\author{James Lloyd-Hughes}
\email{j.lloyd-hughes@warwick.ac.uk}
\affiliation{University of Warwick, Department of Physics, Gibbet Hill Road, Coventry, CV4 7AL, United Kingdom.}

\title{Intertube excitonic coupling in nanotube van der Waals heterostructures}

\begin{document}

\begin{center}
\today  
\end{center}

\begin{abstract}
Excitons dominate the optics of atomically-thin transition metal dichalcogenides and 2D van der Waals heterostructures. Interlayer 2D excitons, with an electron and a hole residing in different layers, form rapidly in heterostructures either via direct charge transfer or via Coulomb interactions that exchange energy between layers. 
Here, we report prominent intertube excitonic effects in quasi-1D van der Waals heterostructures consisting of C/BN/MoS$_2$ core/shell/skin nanotubes.  Remarkably, under pulsed infrared excitation of excitons in the semiconducting CNTs we observed a rapid (sub-picosecond) excitonic response in the visible range from the MoS$_2$ skin, which we attribute to intertube biexcitons mediated by dipole-dipole Coulomb interactions in the coherent regime. On longer ($>100$\,ps) timescales hole transfer from the CNT core to the MoS$_2$ skin further modified the MoS$_2$'s absorption. Our direct demonstration of intertube excitonic interactions and charge transfer in 1D van der Waals heterostructures suggests future applications in infrared and visible optoelectronics using these radial heterojunctions.
\end{abstract}

%%% General paragraph about TMDs/heterostructures/physics?
Atomically-thin semiconductors such as the transition metal dichalcogenides (TMDs) feature electron confinement in two dimensions (2D) and strong Coulomb interactions that produce large exciton binding energies.\cite{Tongay2013, Chernikov2015} TMDs such as MoS$_{2}$ are particularly interesting as semiconductors for optoelectronic devices such as photodiodes and photovoltaic cells,\cite{Tsai2014,Hao2015} however strongly bound excitons do not respond effectively to dc or low-frequency electric fields used in devices.
In van der Waals (vdW) heterostructures dramatic modifications of the electronic and optical properties arise, leading to unique effects like unconventional superconductivity\cite{Cao2018} and excitons confined by Moir\'{e} superlattice potentials.\cite{Tran2019}
The Coulomb interaction between quasiparticles also plays a prominent role in heterostructures, for instance by forming interlayer excitons with electrons and holes located in different layers.\cite{Rivera2015,Jin2018} With their larger polarisability and slower recombination, interlayer excitons provide a route to form free charges more efficiently in TMDs,\cite{Kamban2020} which is desirable to achieve dc current flow in devices. 
 
%%% Quick review of processes in heterostructures
Optical absorption spectroscopy measures the rate at which electrons can transfer between states, either into unbound or bound electron-hole pairs. 
It can be challenging to study interlayer excitons via absorption spectroscopy: the transition rate to form interlayer excitons directly from light is relatively low, as the electron and hole are spatially separated. 
The ultrafast processes required to create interlayer excitons can be studied with sensitive transient absorption spectroscopy methods, by first generating an intralayer exciton that subsequently dissociates across the junction,\cite{Wang2016,Wang2018,Jin2019,Sim2020} or by probing the internal transitions of interlayer excitons.\cite{Merkl2019}
In type II heterojunctions an electron-hole pair generated in one material will separate via charge transfer on ultrafast (sub-picosecond) timescales, creating a long-lived interlayer exciton that straddles the interface. 
Alternatively, F\"{o}rster-type non-radiative dipole-dipole interactions can mediate rapid energy transfer between layers in vdW heterostructures, also on picosecond timescales.\cite{Qian2008,Kozawa2016,Selig2019} 
Later after photoexcitation, once a semiconductor's electrons have cooled by transferring energy to the lattice,\cite{Monti2020} heat transfer can further modify the electronic properties of adjacent layers and thereby change its absorption. 
Disentangling these combined effects of direct charge transfer, Coulomb-mediated energy transfer, and heat transport between layers in vdW heterostructures is thus a significant challenge.

\begin{figure}[!t]
\begin{center}
   \subfloat{\includegraphics[width=1\textwidth]{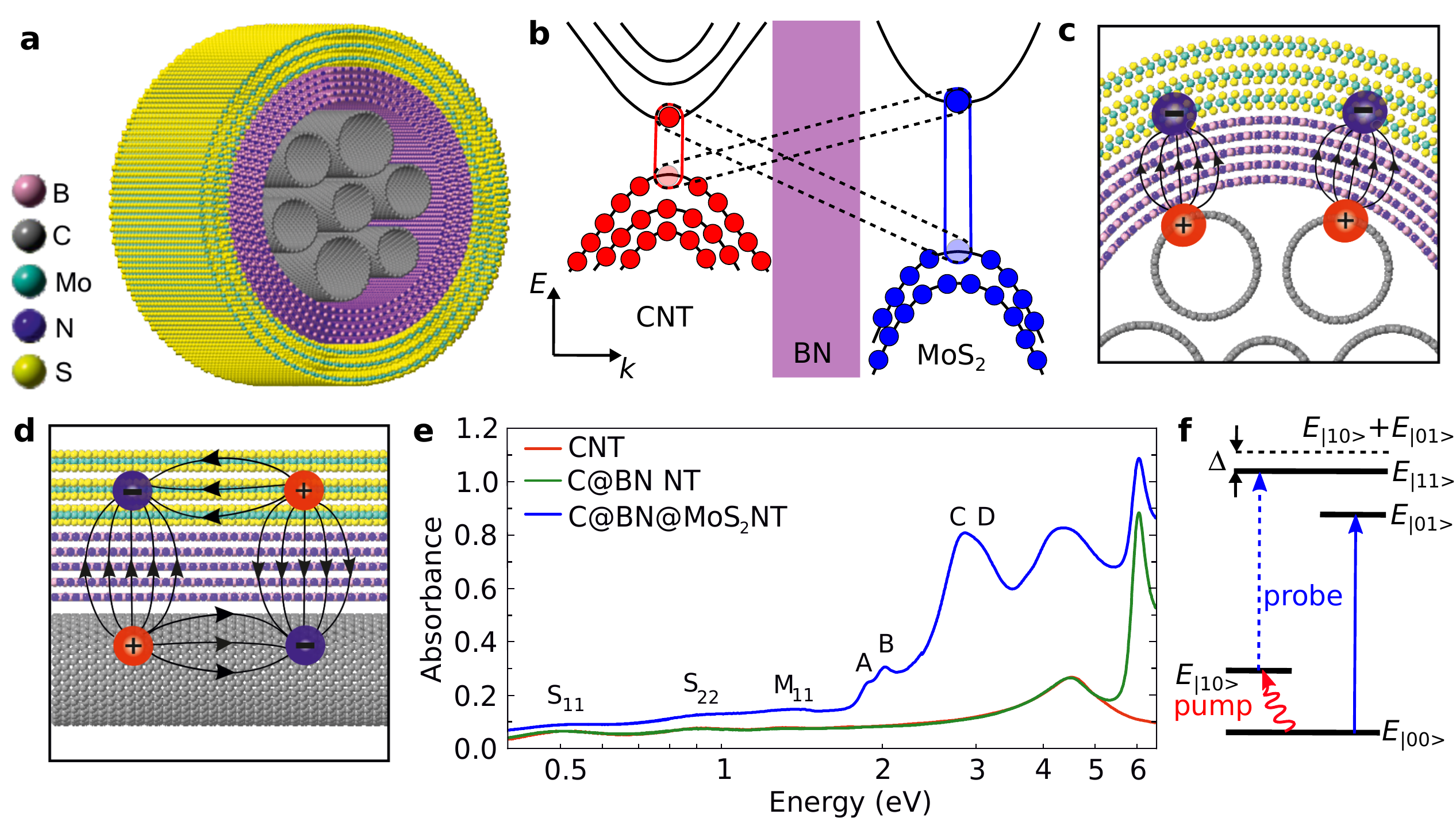}}
\end{center}
\caption{\label{FIG: Concept} \textbf{Electronic and excitonic processes in van der Waals heteronanotubes.} \textbf{a}. Cross-section of a C@BN@MoS$_2$ heteronanotube, comprising CNTs wrapped by BN and MoS$_2$ nanotubes. \textbf{b}. Mixed real-space/single-particle bandstructure schematic of an intertube biexciton. A S$_{11}$ exciton in the CNT (solid red line) and an A exciton in the MoS$_2$ sheath (solid blue line) are bound by attractive Coulomb forces (dashed lines), while the BN (purple) acts as a barrier to direct charge transport. \textbf{c}. A real-space picture of intertube excitons consisting of electrons in the MoS$_2$ and holes in the CNT, viewed along the axis of the heterostructure. \textbf{d}. Intertube biexcitons viewed in a radial cross-section through the heterostructure. \textbf{e}. Optical absorption for pristine CNT films (red), and C@BN (green) and C@BN@MoS$_{2}$ (blue) heteronanotubes. \textbf{f}. In the excitonic picture, with a common ground state energy $E_{|00>}$, the heterostructure has exciton energy levels $E_{|10>}$ for the S$_{11}$ excitons in the CNT and $E_{|01>}$ for an exciton in the MoS$_2$ (e.g.\ the A exciton). The intertube biexciton has an energy level $E_{|11>}$, which is lower in energy than $E_{|01>}+E_{|01>}$ by an amount $\Delta$. Pumping into the S$_{11}$ state at $E_{|10>}$ (red line) lowers the ground-state population, thereby changing the amount of probe light (blue lines) absorbed. }
\end{figure}

Here, we report an investigation of the intertube excitonic response of a nano-coaxial cable, formed by a 1D van der Waals (vdW) heterostructure (Fig.\ \ref{FIG: Concept}a). 
The core consisted of bundles of semiconducting and metallic carbon nanotubes (CNTs), wrapped by insulating BN nanotubes, and with a sheath made of semiconducting MoS$_{2}$ nanotubes (NTs).\cite{Burdanova2020,Xiang2020,Xiang2021} 
The strong excitonic response of the CNTs and the MoS$_2$ NTs, combined with the long-range alignment of the different nanotubes within the heterostructure and large sample areas ($\sim 1$cm$^{2}$), made this an ideal system to study the fundamentals of intertube electronic and excitonic coupling. 
We demonstrate that the Coulomb force creates strong intertube excitonic coupling by comparing the femtosecond transient absorption spectra and dynamics after the initial creation of excitons either in the CNTs or in the MoS$_2$.
%The large tunnel barrier provided by the BN NTs prohibited rapid electron or hole tunneling. 

There are two distinct paradigms by which the light-matter interaction in such a 1D vdW heterostructure may be understood (Fig.\,\ref{FIG: Concept}\textbf{b}). 
Either (i) excitons (solid lines) behave independently of the quasiparticles in the other components of the heterostructure, and the optical properties are an effective medium average of the response of the independent constituents, or (ii) the Coulomb forces between quasiparticles in different layers (dashed lines) create electronic correlations with a unique optical response. 
Simple physical arguments in favour of this second possibility can be advanced. For instance, electrostatics predicts that a free charge in a CNT will be screened by an opposite charge in the sheath (Fig.\,\ref{FIG: Concept}c), forming an intertube exciton. A dipole field in the CNT, i.e.\ an excitonic polarisation, will induce a dipole in the MoS$_{2}$ NT (Fig.\,\ref{FIG: Concept}d), creating an intertube biexciton. 
Dynamically, one may predict that intertube biexcitons and intertube excitons form with different rates following the creation of a coherent excitonic polarisation in the CNT using an infrared pump pulse. 
An intertube biexciton should form rapidly, as it is mediated via the Coulomb force, while intertube excitons require longer to form as charge transfer between the CNT core and MoS$_2$ skin must occur. 
By separating the core and skin with BN nanotubes charge transfer rates can be suppressed (as BN is a good tunnel barrier owing to its wide bandgap), slowing intertube excitonic formation.

Indirect evidence of the importance of intertube interactions comes from the optical absorption, Fig.\ \ref{FIG: Concept}\textbf{e}, obtained for the pristine CNT films, after BN overgrowth, and after MoS$_2$ nanotube growth. As discussed previously, the C@BN@MoS$_2$ heteronanotube films have an optical absorbance from the UV to the THz that can be roughly understood as resulting from the combination of the equilibrium absorbance of the constituent nanotubes in isolation.\cite{Burdanova2020} However, in the infrared range, below the band edge of the MoS$_2$, the C@BN@MoS$_2$ heteronanotube film has an enhanced absorbance in comparison to the C@BN and CNT reference films. One explanation of this effect is that Coulomb correlations between the MoS$_2$ and the CNTs enhances the excitonic absorption of the S$_{11}$, S$_{22}$, M$_{11}$ excitons in the CNTs, much in the way that electron-hole pairs boost the absorption strength near the band edge of a direct gap semiconductor. Alternatively, the MoS$_2$ growth may have modified the effective dielectric function, increasing the amount of reflection loss, or a fraction of the MoS$_2$ nanotubes may be metallic\cite{Milosevic2007} and hence absorb in the infrared. 
Photoluminescence (PL) spectroscopy is also sensitive to intertube interactions: for instance the A exciton PL efficiency was found to be enhanced for BN@MoS$_2$ NTs with respect to that of C@BN@MoS$_2$ NTs.\cite{Liu2020}

In order to establish uniquely whether intertube Coulomb interactions modify the ground state electronic and the optical properties of 1D vdW heterostructures, we introduce a scheme based on multi-colour pump-probe spectroscopy (Fig.\ \ref{FIG: Concept}\textbf{f}). 
A heterostructure with Coulomb-coupled states has excitonic energy levels $E_{mn}$, where $m,n=0$ denotes the ground state, $|00>$, $m=1$ denotes an S$_{11}$ exciton in a semiconducting CNT, and $n=1$ labels an A or B exciton in the MoS$_2$. 
By pumping at an energy $E_{1}=E_{|10>}-E_{|00>}$ in the infrared, a fraction of the system is moved from the ground state $|00>$ to the $|10>$ state. 
If the heterostructure has a common Coulomb-coupled ground state then an optical probe at $E_2=E_{|01>}-E_{|00>}$ exhibits a reduced absorption as a result of the lower ground state population. 
In contrast, if the MoS$_{2}$ NT and CNTs are electronically independent, and there are no Coulomb correlations between NTs, then pumping at $E_1$ will not induce an absorption change at $E_2$: the ground state population of the MoS$_2$ NT will not be modified. 
By checking whether or not infrared excitation of BN@MoS$_2$ NTs modifies the visible absorption, one can further test if MoS$_2$ NTs have any inherent absorption processes below the A exciton.
 
\begin{figure}[!t]
\begin{center}
   \subfloat{\includegraphics[width=1\textwidth]{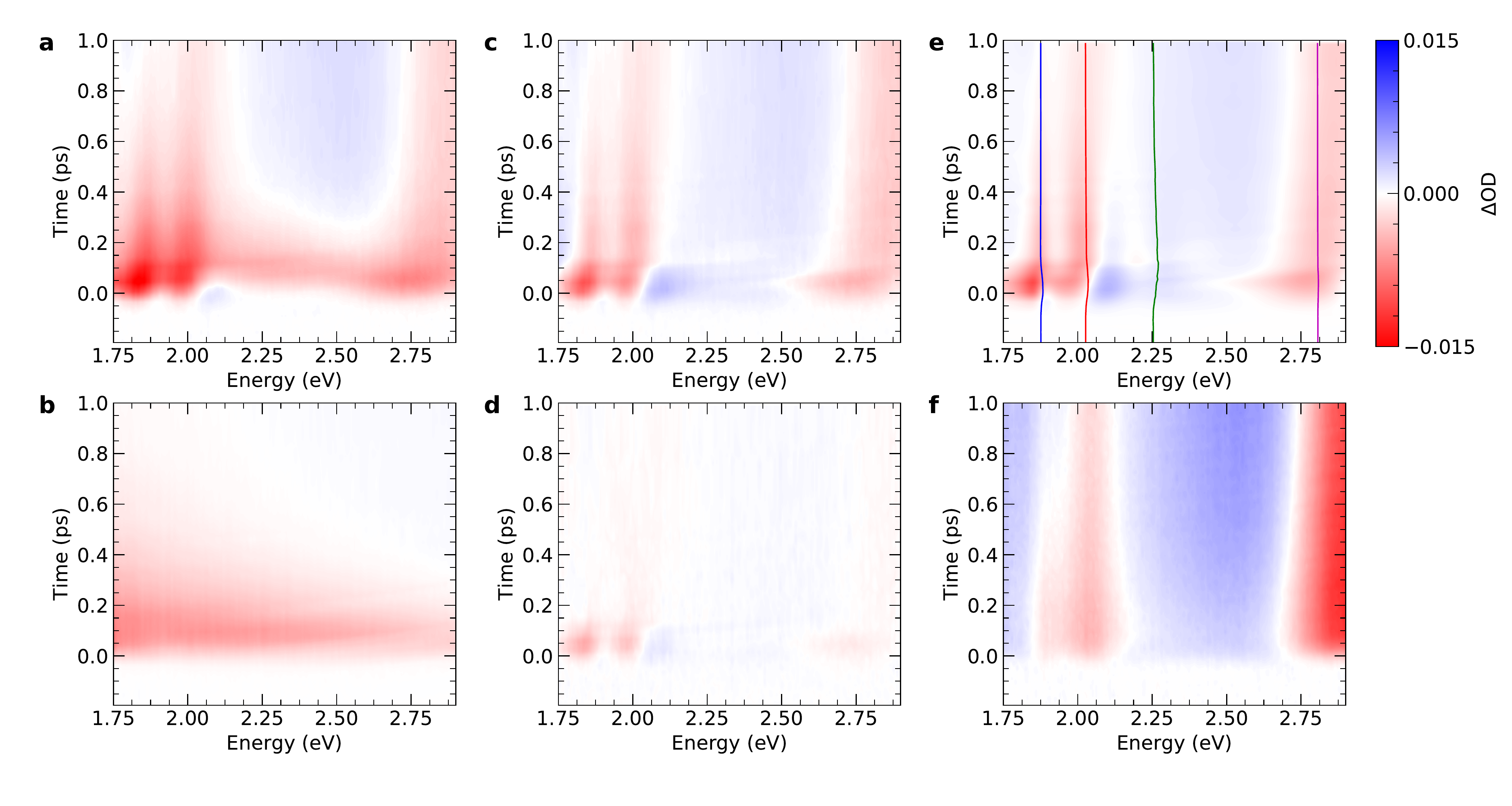}}
\end{center}
\caption{\label{FIG: TransientAbsorption} Transient absorption spectra at a range of pump-probe delay times for \textbf{a}. C@BN@MoS$_{2}$ NTs and \textbf{b}. C@BN NTs, under excitation with a 0.62\,eV, 200 $\mu$Jcm$^{-2}$, 35\,fs pump pulse. \textbf{c}. Data from \textbf{a} after subtracting the CNT response in \textbf{b}. \textbf{d}. As \textbf{c}, but for data obtained at pump fluence 70\,$\mu$Jcm$^{-2}$. \textbf{e} Multi-oscillator fit to the optical density under photoexcitation from panel \textbf{c}. Vertical lines show the fitted oscillator energies versus time. \textbf{f}. Excitation with pump photons at 3.0\,eV produces changes dominated by the A, B and interband absorption of the MoS$_2$. The same colour map (top right) was used for each panel.}
\end{figure}

Ultrafast absorption spectroscopy experiments on thin films of C, C@BN, C@BN@MoS$_2$ and BN@MoS$_2$ nanotubes were performed, pumping at 0.6\,eV (close to $E_1$) or at 3.0\,eV, with a white-light probe that covered the A (1.88\,eV) and B (2.02\,eV) excitons and interband absorption region of MoS$_2$. 
The maps of the differential change in optical density, $\Delta$OD, in Fig.\ \ref{FIG: TransientAbsorption}a demonstrate that the MoS$_2$ NTs' optical response is modified by the 0.6\,eV pump: for instance the A and B excitonic absorption peaks are suppressed (red areas are reduced optical density; higher transmission). 
As the pump photon energies (0.6\,eV) were only a third of the band edge of MoS$_2$ (1.88\,eV), and a BN@MoS$_2$ sample prepared with the CNTs removed showed no discernible $\Delta$OD (Supplemental Figs.\ S1-2), processes involving single-photon absorption by the MoS$_2$ NTs (e.g.\ band-tail absorption; free-carrier absorption) can be ruled out. 
Non-linear absorption processes such as three-photon absorption can also be eliminated by the lack of response from BN@MoS$_2$ NTs, further confirmed by the linearity of $\Delta$OD with pump fluence (Supplemental Fig.\ S3). 
The modification of the excitonic response of the MoS$_2$ can therefore be uniquely identified as resulting from coupling between the S$_{11}$ excitons generated in the CNTs and the electronic states of the MoS$_2$.

To further elucidate the dynamic response of the C@BN@MoS$_2$ heteronanotubes, we performed reference experiments with a C@BN NT film without MoS$_2$ NTs. 
Figure \ref{FIG: TransientAbsorption}\textbf{b} shows a broadband, weaker response from the higher lying excitonic transitions in the CNTs (e.g.\ S$_{33}$, S$_{44}$ and their continua). 
This transient response from the CNTs decays with lifetimes below 220\,fs (Supplemental Fig.\ S4). 
To examine the optical response of the MoS$_2$ NTs without this contribution we subtracted the reference's $\Delta$OD data set from that of the heteronanotubes, yielding Fig.\ \ref{FIG: TransientAbsorption}\textbf{c}, where features near zero pump-probe delay can be seen more clearly. 
Using data obtained at lower fluence ($70$\,$\mu$Jcm$^{-2}$), the same procedure yields Fig.\ \ref{FIG: TransientAbsorption}\textbf{d}. 
At early times and lower fluence the response from the MoS$_2$ NTs persists for only 100\,fs, with peaks in $\Delta$OD near the A and B excitons and C peak (2.81\,eV). 
The $\Delta$OD spectra at higher fluence are similar for the first 100\,fs, whereapon a more persistent $\Delta$OD signal with a distinct spectral response appears. 

%The dynamical change in $\Delta$OD can be understood with 

Physically, the dynamics at early time result from the strong electric field of the IR pump pulse: this creates a coherent S$_{11}$ excitonic polarisation in the CNTs, which then induces a polarisation response from the MoS$_2$ nanotubes via the Coulomb interaction, forming transient intertube biexcitonic states, as in Fig.\ \ref{FIG: Concept}d. 
The timescales for the transient rise and decay of this intertube exciton-exciton coupling were examined by fitting the experimental transmittance spectra at each pump-probe delay with a modelled transmittance calculated from a dielectric function built from multiple oscillators. 
To cover the experimental range of the transient absorption spectra, four oscillators were adopted (A and B excitons, interband transitions and the C peak), and their centre energy, oscillator strength and linewidth were allowed to vary with delay. 
An example of the $\Delta$OD resulting from this fit is reported in Figure \ref{FIG: TransientAbsorption}e, while the centre energy and strength of the A exciton are reported in Figure \ref{FIG: Fits} (full details of the fits and parameters are provided in the Supplemental Figs.\ S5-7). The dynamics following ultrafast excitation at 3.0\,eV were also measured (Fig.\ \ref{FIG: TransientAbsorption}f) and fitted: the extremely fast formation of band-edge excitonic response was below 100\,fs, in line with recent results on monolayer MoS$_2$, and attributed to rapid exciton cooling mediated by the strong exciton-phonon interaction.\cite{Trovatello2020} 
\begin{figure}[tb]
\begin{center}
   \subfloat{\includegraphics[width=1\textwidth]{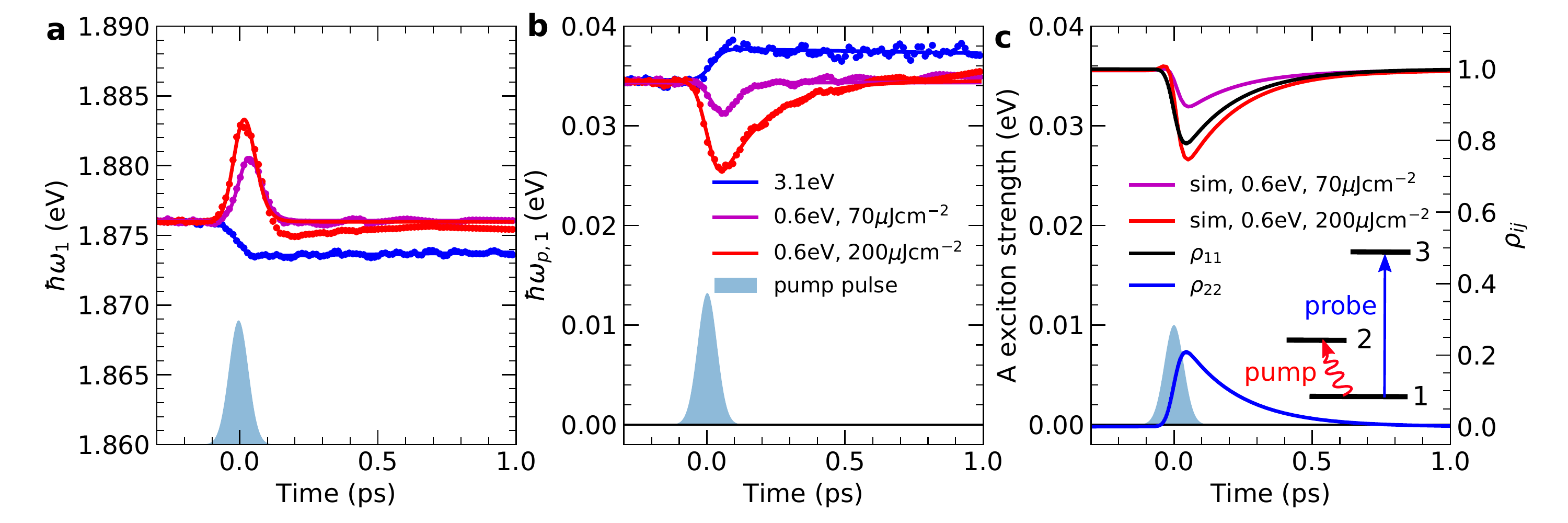}}
\end{center}
\caption{\label{FIG: Fits} Time-dependent dynamics of the A exciton. \textbf{a}. The A exciton energy, $\hbar\omega_1$, blueshifts during the coherent regime only, with a larger blueshift for 200\,$\mu$Jcm$^{-2}$ (red points) than 70\,$\mu$Jcm$^{-2}$ (purple points). UV excitation produces a quasi-permanent redshift (blue points). Solid lines are fits using a Gaussian IR pulse envelope (shaded area, $\sigma=35$\,fs) and an exponential decay. \textbf{b}. As in \textbf{a}, but for the A exciton's oscillator strength, $\hbar\omega_{p,1}$. \textbf{c} Bloch equation simulations of the ground state and CNT exciton populations, $\rho_{11}$ (blue) and $\rho_{22}$ (black), along with the simulated strength of the A exciton response at the same fluences as used in experiment (red and purple lines).}
\end{figure} 
 
Under IR excitation the A exciton's centre energy undergoes a dynamic blueshift, evident in Figure \ref{FIG: Fits}a, which rises and decays with the IR laser pulse (shaded area, Gaussian pulse with $\sigma=35$\,fs). 
The peak blueshift increased with fluence. 
Similar blueshifts were seen for the B exciton and the interband oscillator at 2.25\,eV under IR excitation (Supplemental Fig.\ S7), whereas direct excitation of the MoS$_2$ NTs at 3.0\,eV produced a redshift in the A exciton energy as a result of bandgap renormalisation.\cite{Zhu2017}
Within the coherent regime (during the IR pulse), the blueshift may be understood as an optical Stark shift of the exciton energies in the MoS$_2$.
Alternatively, with a finite population in state $E_{10}$ there is an excited state absorption pathway (dashed line in Fig.\ \ref{FIG: Concept}f) at an energy $E_{|11>}-E_{|10>}=E_{|01>}-E_{|00>}-\Delta$, i.e.\ an energy $\Delta$ below the excitonic resonances of the MoS$_2$ NTs.
This excited state absorption could therefore counteract the ground state bleach of the $E_{|00>}$ to $E_{|01>}$ transition, if $\Delta$ is comparable to the linewidth of the excitonic absorption. 
As the magnitudes of the blueshifts are small compared to their linewidths we cannot conclusively identify their origin.

The A oscillator strength (Figure \ref{FIG: Fits}b) lowered during the IR pump pulse, directly showing that fewer electrons were available to undergo the transitions that form A excitons. 
At times after the IR pump pulse the intertube excitonic response recovered rapidly towards equilibrium, with a time constant $\tau$ that was determined from fits to the dynamics (solid lines in Figure \ref{FIG: Fits}).
At lower IR fluence we found $\tau=43$\,fs, suggesting the intertube excitonic response persisted essentially only within the coherent regime, while at higher IR fluence the intertube excitonic response was longer lived ($\tau=166$\,fs).
After the IR drive field has finished, the coherent S$_{11}$ polarisation decoheres at a rate set by both exciton momentum scattering and exciton population transfer processes. 
For example, optically-bright S$_{11}$ singlet excitons can relax to lower energy optically-dark triplet excitons. 
Alternatively, they can transfer their energy via nonradiative near-field coupling to another nearby nanotube with a finite density of states at the same energy,\cite{Mehlenbacher2015} such as a metallic CNT. 
Regardless of the mechanism by which the coherent S$_{11}$ excitonic polarisation is removed, the induced excitonic response of the MoS$_2$ is lowered when the S$_{11}$ polarisation weakens. 
Decoherence of the excitonic polarisation component in the MoS$_2$ can also cause the destruction of intertube correlations. 
Hence the intertube excitonic contribution decays on timescales comparable to, or faster than, the timescale for recovery of the C@BN reference sample's interband response, which was 220\,fs and faster (Supplemental Fig.\ S4). The excitonic response of A and B excitons directly created in the MoS$_2$ NTs under 3.0\,eV excitation recovered more slowly (Fig.\ \ref{FIG: TransientAbsorption}f).

%When charges have transferred from semiconducting tubes to metallic tubes, the transient absorption will have returned to its equilibrium level, but the electrons and holes in the metallic tubes still have to relax further energetically.

A full theoretical treatment of intertube excitonic coupling in the heterostructure requires the self-consistent inclusion of the semiconductor physics, for instance via the semi-classical semiconductor Bloch equations, along with consideration of the electromagnetism by solving Maxwell's equations. 
The current state-of-the-art in this area includes studies of superradiant effects in coupled 2D TMD layers \cite{Stevens2018} and the rapid ultrafast formation dynamics of excitons in MoS$_2$ monolayers after non-resonant visible excitation.\cite{Trovatello2020} 
Here we provide two simplified theoretical models that allow intuitive insights into intertube excitonic effects. 
Firstly, a pair of coupled classical oscillators can represent the system: a first oscillator represents the excitonic polarisation of an S$_{11}$ exciton in a CNT, the second oscillator denotes an A exciton in a MoS$_2$ NT, while a third spring models the Coulomb dipole-dipole coupling between excitons. 
When an impulse produces an excitonic polarisation in the CNT oscillator (represented by the extension of the first spring from equilibrium) then a damped oscillatory response from the MoS$_2$ oscillator is produced, with a spectral maximum close to the A exciton energy (Supplemental Information S8). 

Alternatively, a second theoretical model is to solve the Bloch equations in the excitonic basis for the lowest three levels in Fig.\ \ref{FIG: Concept}\textbf{f}, as described in the Supplemental Information. 
This approach improves on the classical model by including quantum effects such as finite level occupancy and dipole matrix elements, and includes electromagnetism self-consistently, but ignores many-body effects (e.g.\ bandgap renormalisation) and the optical Stark effect, and includes dephasing and recombination only phenomenologically. 
Figure \ref{FIG: Fits}c illustrates the output from this simulation, with population decay times $\tau_d=200$\,fs and $60$\,ps for S$_{11}$ excitons (level 2) and MoS$_2$ A excitons (level 3) respectively. 
During excitation of the $1\rightarrow2$ transition (shaded area), the rapid dephasing of the coherent excitonic response (represented by the components $\rho_{12}$ and $\rho_{21}$ of the density matrix) produced an S$_{11}$ exciton population, $\rho_{22}$.
The ground state population, $\rho_{11}$, and $\rho_{22}$ both returned towards equilibrium at a rate set by the fast population decay $\tau_d$. Further, a full simulation of the pump-probe experiment was obtained using multiple Bloch simulations including a probe pulse resonant to the $1\leftrightarrow3$ transition and arriving at varying time delays. The dynamics of the pump-induced change in A oscillator strength from this model (red and purple lines in Figure \ref{FIG: Fits}c) track the decay in S$_{11}$ population, and are in excellent accord with the experimental data (Figure \ref{FIG: Fits}b). This theoretical description provides further evidence that infrared excitation of excitons in the CNT core can yield strong modifications in the visible response of the MoS$_2$ NT sheath.

\begin{figure}[tb!]
\begin{center}
   \subfloat{\includegraphics[width=1\textwidth]{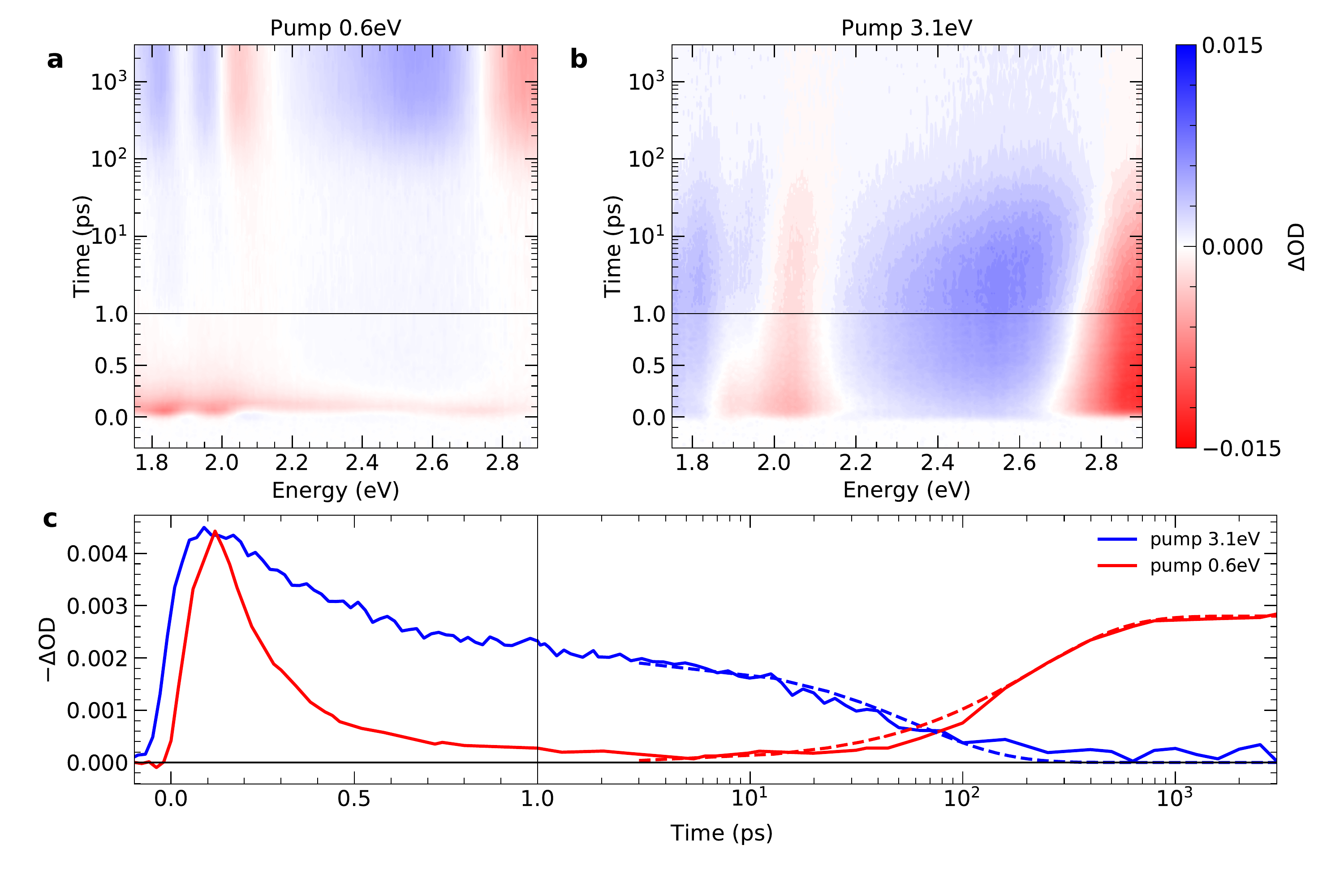}}
\end{center}
\caption{\label{FIG: Pump-probe-later-time} Time evolution of the pump induced change in OD for the C@BN@MoS$_2$ heteronanotube film at later pump-probe delay times. \textbf{a}. Following IR excitation at 70\,$\mu$Jcm$^{-2}$, the coherent intertube excitonic effects disappear within 1\,ps, followed by a slower increase in the response from the MoS$_2$, which then does not decay within the 3\,ns time window of the experiment. \textbf{b}. Under direct (UV) excitation at 200\,$\mu$Jcm$^{-2}$ the transient OD recovers monotonically towards equilibrium. \textbf{c}. Slices at constant energy (averaged over 2.02-2.04\,eV around the B exciton) reveal the differing dynamics at later times under IR excitation. Simple fits (dashed lines) allow indicative timescales to be extracted: an exponential decay (of the form $y(t)=a e^{-t/\tau_d}$, with $\tau_d=60$\,ps) for direct excitation of the MoS$_2$ at 3.0\,eV, or an exponential rise (of the form $y(t)=a(1-e^{-t/\tau_r})$, with $\tau_r=220$\,ps) for excitation at 0.6\,eV.}
\end{figure} 

Finally, we report the transient absorption dynamics of the heteronanotubes at later times after excitation. Under infrared excitation, the initial ultrafast response (discussed above) disappears within 1\,ps, but is followed by a gradual enhancement of transient absorption at later times, evident in Figure \ref{FIG: Pump-probe-later-time}a. This change is then long-lived, remaining constant until the end of the 3\,ns experimental time window. Similar features were observed at different pump fluences, and were not observed on BN@MoS$_2$ or C@BN reference samples. This transient response indicates that a slower process, such as heat transport or charge transfer from the CNTs to the MoS$_2$ NTs, has modified the absorption associated with the A and B exciton and interband absorption features of the MoS$_2$ NTs. A photothermal explanation can be ruled out by examining the dynamics under UV excitation: under comparable pulse fluences UV and IR excitation would ultimately heat the NTs in a similar way and would thus be expected to produce the same long-lived response. However, excitation of the MoS$_2$ NTs at 3.0\,eV led to a $\Delta$OD that reduced monotonically with time, with a characteristic lifetime of 60\,ps (Fig.\ \ref{FIG: Pump-probe-later-time}b-c) that was much faster than the long-lived feature under IR excitation, which persisted for $\gg1$\,ns. 

Instead, the formation of indirect intertube excitons via charge transfer can account for the slow response under IR excitation. 
Indeed the characteristic timescale of 220\,ps for its rise (Fig.\ \ref{FIG: Pump-probe-later-time}c) is consistent with our estimates of hole tunneling times from the CNTs to the MoS$_2$: using a model of quantum tunneling through the BN barrier\cite{Yadav2020} we estimated hole tunneling times of $100$\,ps-$1$\,ns  (Supplemental Fig.\ S9). 
While the indirect excitons thus created do not directly absorb, the transfer of quasiparticles from one layer to another modifies the absorption rate in the same way as discussed for the ultrafast intertube response.

%\section*{Discussion}
In summary, the results presented in this work show that intertube coupling plays a pivotal role in the optical response of quasi-1D van der Waals heterostructures. 
Infrared excitation of the carbon nanotube cores created a response from the MoS$_2$ sheath on two different timescales. The initial ultrafast response at early times (around 100\,fs) was dominated by intertube excitonic coupling mediated by the Coulomb interaction, while charge transfer via quantum tunneling at later times (around 100\,ps) produced indirect intertube excitons.
The results demonstrate that the visible and UV optical properties of one component of a heterostructure (here the MoS$_{2}$ NTs) can be manipulated by selective excitation of another constituent (the CNTs). 
Therefore, efficient intertube coupling and the effective control of light at different wavelengths can be readily achieved in heteronanotubes.

\section*{Acknowledgements}
MGB acknowledges support from the Global Education Program (Russia) for a PhD scholarship. The UK authors acknowledge the EPSRC (UK) for funding under grant EP/N010825/1. Part of this work was supported by JSPS KAKENHI (grant numbers JP18H05329, JP19H02543, JP20H00220, and JP20KK0114) and by JST, CREST grant number JPMJCR20B5, Japan.
  
\providecommand{\latin}[1]{#1}
\makeatletter
\providecommand{\doi}
  {\begingroup\let\do\@makeother\dospecials
  \catcode`\{=1 \catcode`\}=2 \doi@aux}
\providecommand{\doi@aux}[1]{\endgroup\texttt{#1}}
\makeatother
\providecommand*\mcitethebibliography{\thebibliography}
\csname @ifundefined\endcsname{endmcitethebibliography}
  {\let\endmcitethebibliography\endthebibliography}{}

\end{document}